\documentclass{pas}
\usepackage{amsmath}
\usepackage{natbib}

\newcommand{\aap}{A\&A}
\newcommand{\apjl}{ApJ}

\newcommand{\apjs}{ApJS}

\usepackage{multirow}
\usepackage{graphicx}
\usepackage{appendix}
\usepackage{etex}
\usepackage{gensymb}
\usepackage{amsmath,siunitx}
\usepackage [english]{babel}
\usepackage [autostyle, english = american]{csquotes}
\MakeOuterQuote{"}

\begin{document}

\lefttitle{Tracing the evolutionary pathway of Polar Ring Galaxies in the cases of NGC 3718, NGC 2685, and NGC 4262}
\righttitle{Akhil et al. 2025}

\jnlPage{1}{17}
\jnlDoiYr{2025}
\doival{PAS-2024-0112}

\articletitt{Research Paper}

\title{Connecting the dots: \\ Tracing the evolutionary pathway of Polar Ring Galaxies in the cases of NGC 3718, NGC 2685, and NGC 4262}

\author{\sn{Krishna R.} \gn{Akhil}, \sn{Sreeja} \gn{S Kartha} , \sn{Ujjwal} \gn{Krishnan}, \sn{Blesson} \gn{Mathew}, \sn{Robin} \gn{Thomas}, \sn{Shankar} \gn{Ray}, and \sn{Ashish} \gn{Devaraj}  }

\affil{Department of Physics and Electronics, CHRIST (Deemed to be University), Bangalore, India}

\corresp{Sreeja S. Kartha: sreeja.kartha@christuniversity.in, \\ Akhil Krishna R.: akhil.r@res.christuniversity.in}

%\authorrunning{Akhil et al. 2024 }
%\titlerunning{Polar Ring Galaxy NGC 4262 }
\citeauth{Akhil et al. 2025, Connecting the dots: Tracing the evolutionary pathway of Polar Ring Galaxies in the cases of NGC 3718, NGC 2685, and NGC 4262. {\it Publications of the Astronomical Society of Australia}}

\history{(Received 08-Jun-2024; accepted 05-Mar-2025)}

\begin{abstract}

Polar Ring Galaxies (PRGs) are a unique class of galaxies characterised by a ring of gas and stars orbiting nearly orthogonal to the main body. This study delves into the evolutionary trajectory of PRGs using the exemplary trio of NGC 3718, NGC 2685, and NGC 4262. We investigate the distinct features of PRGs by analysing their ring and host components to reveal their unique characteristics through Spectral Energy Distribution (SED) fitting. Using CIGALE, we performed SED fitting to independently analyse the ring and host spatially resolved regions, marking the first decomposed SED analysis for PRGs, which examines stellar populations using high-resolution observations from \textit{AstroSat} UVIT at a resolved scale. The UV-optical surface profiles provide an initial idea that distinct patterns in the galaxies, with differences in FUV and NUV, suggest three distinct stages of ring evolution in the selected galaxies. The study of resolved-scale stellar regions reveals that the ring regions are generally younger than their host galaxies, with the age disparity progressively decreasing along the evolutionary sequence from NGC 3718 to NGC 4262. Star formation rates (SFR) also exhibit a consistent pattern, with higher SFR in the ring of NGC 3718 compared to the others, and a progressive decrease through NGC 2685 and NGC 4262.  Finally, the representation of the galaxies in the HI gas fraction versus the NUV--$\textnormal r$ plane supports the idea that they are in three different evolutionary stages of PRG evolution, with NGC 3718 in the initial stage, NGC 2685 in the intermediate stage, and NGC 4262 representing the final stage. This study concludes that PRGs undergo various evolutionary stages, as evidenced by the observed features in the ring and host components. NGC 3718, NGC 2685, and NGC 4262 represent different stages of this evolution, highlighting the dynamic nature of PRGs and emphasising the importance of studying their evolutionary processes to gain insights into galactic formation and evolution.

\end{abstract}


\maketitle
\section{Introduction}
\cite{whitmore1990} defined Polar Ring Galaxies (PRGs) as unique galaxies belonging to the lenticular type (S0), characterised by a ring or disc orbiting in a nearly polar plane relative to the main galaxy. Further studies mentioned that the host galaxies are often gas-poor S0 or elliptical (E) types, while the ring consists of gas, dust, and stars \citep{reshkinov1997,finkelman2012}. This definition, however, has evolved with recent observations indicating that both red and blue star-forming host galaxies can exhibit orthogonal rings \citep{2015Reshetnikovandcombes} and that these rings are not always star-forming \citep{2023Degwallaby}. The major scenarios proposed for the formation of PRGs are major galaxy mergers \citep{1998Bekki,2003Bournaud_Combes}, tidal accretion of mass from a nearby companion \citep{reshkinov1997,2003Bournaud_Combes}, and cold gas accretion from the intergalactic medium \citep{2008ApJ...689..678B}.\par
According to \cite{1998Bekki}, PRGs have the capability to transform from a narrow ring structure to peculiar double rings or transient annular rings. \citet{2003Bournaud_Combes} highlight that the ring structure in PRGs can remain stable even after 8 Gyrs in both merging and accretion scenarios. Also, it is {seen in simulations} that the majority of polar rings exhibit a warp, which typically emerges after 1 to 3 Gyrs of formation. Studies on the stability of self-gravitating polar rings suggest the likelihood of observing warps \citep{1994Arnaboldisparke,2009Sparke}. Warp signatures are also discernible in the shape profiles of simulated rings, which is in line with the anticipated behaviour of self-gravitating polar rings \citep{2024Smirnov}. Also, the properties of PRGs are more similar to those of early-type galaxies than late-type galaxies \citep{whitmore1990,2024akr_2}. Therefore, exploring the captivating features of the evolution of PRGs will greatly enhance our understanding of the total galaxy evolution framework. \par 

The ring structure of PRGs is typically much bluer than the host component, indicating recent star formation activity \citep{2022Mosenkov,2024akr}. In this study, we obtain the spectral energy distribution (SED) of the galaxies and their subcomponents to gain further insight into various parameters, such as star formation rate, stellar mass, and age. SED modelling techniques play a crucial role in exploring these galaxies, enabling us to extract valuable insights about their distinct emission sources \citep{1989Cardelli_SED,2005Burgarella_SED}. In this paper, we employ the SED fitting code-named 'Code Investigating GALaxy Emission' \citep[CIGALE; ][]{2005Burgarella_SED,2009Noll_cigale,2019Boquien} to analyse a multi-wavelength dataset of a selected sample of PRGs. {In this study, we aim to investigate the resolved-scale stellar populations in the host and ring components of PRGs separately across UV, optical, and mid-IR wavelengths. We analyse their SEDs separately for regions belonging to the ring and the host. This separation is challenging, as overlapping components, particularly at certain viewing angles, can affect the photometric data of host galaxies. To assess how these variations and flux measurements influence the SED, we analyse spatially resolved regions within the galaxy, distinguishing between the host, ring, and their overlapping areas.} This approach enables us to determine the physical properties and star formation histories of the ring and host components separately, offering key revelations into the evolutionary pathways of PRGs. \par

To comprehensively investigate and gain a better understanding of the star-forming activities within the ring components, as well as the evolutionary scenarios of PRGs, we use UV data from the Ultra-Violet Imaging Telescope (UVIT). The UVIT onboard {\it AstroSat} possesses the capacity to provide detailed insights into the star-forming properties of galaxies, as evidenced by the recent studies \citep{2018george,2022UK,2023Ashish,2024Mayya,2024Sundar,2024Leahy,2024Ujjwal,2024Robinjigsaw,2024robin,2025Shashank}. Using GALEX data, previous investigations \citep{marino2009,ordernes2016A&A...585A.156O} have explored the star-forming regions and their characteristics in PRGs. However, these studies were constrained by the lack of deep and multi-band photometric data in the UV regime. The multiple filters in the UVIT instrument are useful for better fitting galaxy SEDs. Also, the utilisation of UV images from UVIT to identify and study star-forming regions in the ring is anticipated to provide a more detailed understanding of the evolution of PRGs. \par
Details regarding the selected sample of galaxies are presented in Section. \ref{sec:samp}. In Section \ref{sec:obs}, the observations and data reduction procedures are outlined. Subsequently, in Section \ref{sec:anal_res}, the analysis is conducted, and the corresponding results are presented. Finally, we discuss and summarise the results in Sections \ref{sec:disc} and \ref{sec:conclu}, respectively.

\begin{figure*}[hbt!]
    \begin{center}
        \includegraphics[width = 1.55\columnwidth]{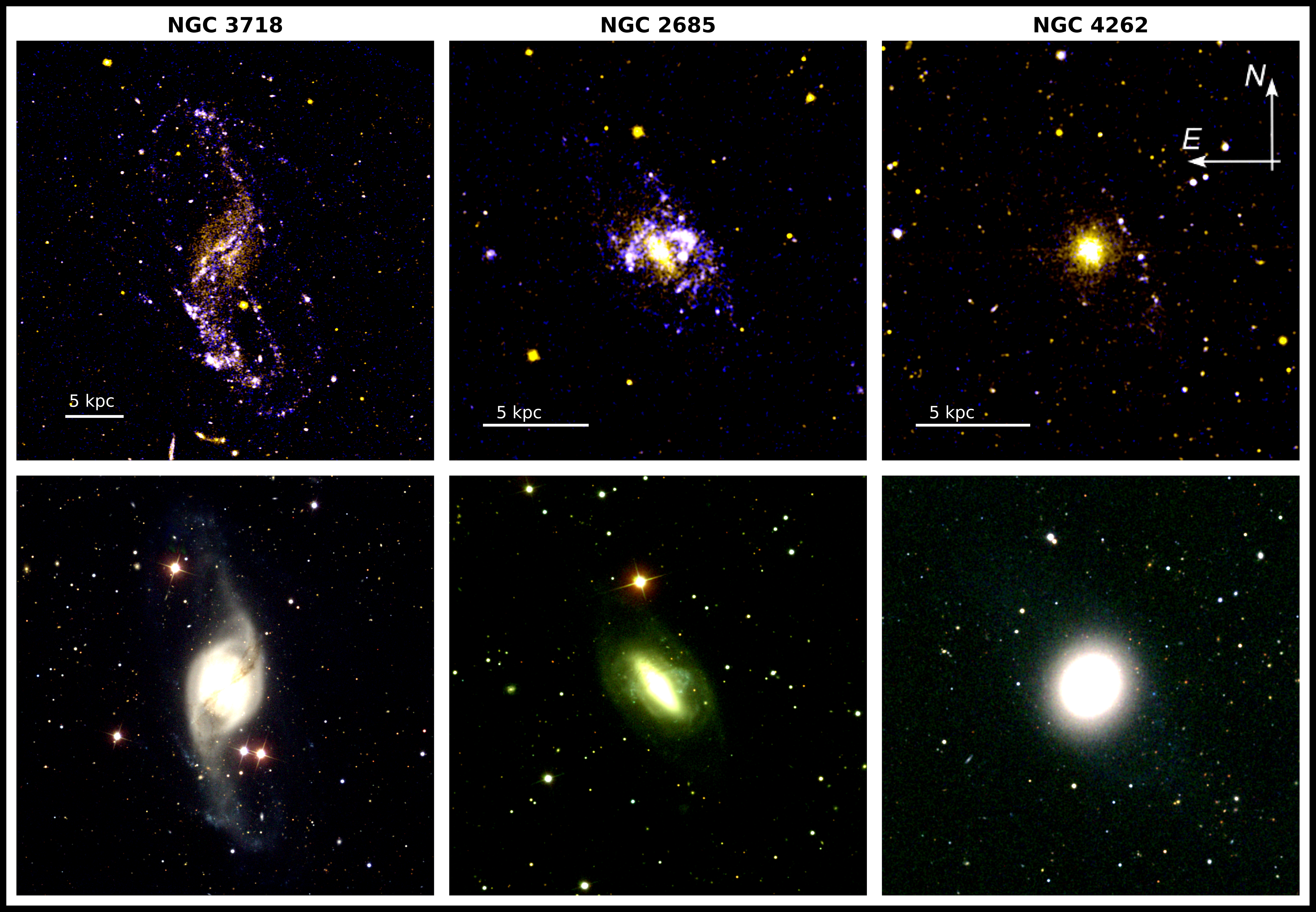}
        \caption{The UV and Optical colour composites of the selected three PRGs are shown here. The upper panel showcases the UV colour composite, where the blue and yellow colours correspond to the images in UVIT FUV and NUV filters, respectively. The physical scale bar of 5 kpc is shown in the bottom left corner. The lower panel exhibits the optical colour composite, with blue, green, and red colours representing the galaxy images in SDSS {\it g, r,} and {\it i} bands, respectively.}
        \label{fig:color}
    \end{center}
\end{figure*}

\section{A brief synopsis of the sample of PRGs used for this study.}
\label{sec:samp}
{This paper aims to study the evolution of PRGs through observational effects in the ring structure.} From observational data, galaxies such as NGC~3808 \citep{1996Reshetnikov3808,ordernes2016A&A...585A.156O} and NGC~6286 \citep{2004Shalyapina6286} are identified as being in the ring formation process through interactions. To find the observational features in the ring evolution of PRGs, we selected a sample of PRGs located within 20 Mpc. The availability of UVIT multiple filter observations will be used to explore the energy distributions of the ring and host separately for galaxies. Moreover, the availability of UVIT observations is crucial for investigating the star-forming regions in the PRGs on a resolved scale. Furthermore, we require nearby well-studied PRGs; hence, we obtained three highly suitable galaxies for this proposed study: NGC 3718, NGC 2685, and NGC 4262. The {physical properties} of the {three selected galaxies} are listed in Table \ref{tab:basic}.\par
The discussion on the formation mechanism of the ring structure of NGC 3718 began with \cite{1985Schwarz}, and it remains unclear whether it is due to merging or accretion. \cite{2015markakis} have explored the possibility of a merger. Recently, \cite{2024Chandan} proposed that the asymmetric arm structures in NGC 3718 could be due to the passage of a low-mass companion galaxy. These studies indicate that subsequent merging, after the possible accretion of gas into a polar orbit, could also play a role in forming NGC 3718. However, the symmetry of NGC 3718 and the weak bridge between NGC 3718 and NGC 3729 suggest the possibility of accretion events. \cite{2004Pott}, \cite{2004Jozsa}, and \cite{2005Krips} also discussed the gas distributions and symmetry of galaxies in this context.

In the case of NGC 2685, the formation scenario of the ring is confirmed to involve past accretion events based on the amount of molecular and atomic gas content, as discussed by \cite{1998silchenko}, \cite{2014silchenkoandmoiseev} and \cite{2002Schinnerer}. Similarly, in the case of NGC 4262, \cite{2024akr_2} discussed the possibility of recent accretion events based on the properties of the globular cluster systems.

\begin{table}[hbt!]
    \caption{{The fundamental physical characteristics of the three galaxies. The references for each property are given as footnotes for the table. The parameters listed include right ascension and declination, distance, apparent diameter in kiloparsecs, position angle, total V-band apparent magnitude, NUV magnitude, logarithmic HI total mass, logarithmic stellar mass, and logarithmic star formation rate.}}
    \scalebox{0.85}{
    \begin{tabular}{llll}
        \hline
        {Parameters} & {NGC 3718} & {NGC 2685} & {NGC 4262} \\ \hline
        RA$^{1}$ (deg)                                & 173.145 &133.894 & 184.877 \\
        Dec$^{1}$ (deg)                                & 53.067 &58.734 &14.877 \\
        Distance$^{2}$  (Mpc)                  & 18.67  &16.78 & 15.5 \\
        Apparent diameter$^{3}$    (kpc)         &46.93&  18.98& 10.94\\
        PA $^{6}$ (deg)            &11.4&38& 145 \\
        V$_{T}$ $^{6}$ (mag)                   &10.69&11.35&11.46  \\ 
        NUV $^{7}$ (mag)						&14.15&15.08&16.12  \\
        Log HI total mass$^{4}$ ($\textup{M}_\odot$)  &10.5& 9.2 & 8.6 \\
        Log stellar mass$^{5}$ ($\textup{M}_\odot$) &10.48&10.13&10.06	\\
        Log SFR$^{2}$ ($\textup{M}_\odot$yr$^{-1}$)                  &-0.66& -0.82	&-1.97 \\
        \hline
    \end{tabular}}
    \label{tab:basic}
    \footnotesize{References 1. \cite{2015Alam}, 2. \cite{2018dalya_dist}, 3. \cite{1991morph_and_redshift} 4. \cite{2019DeVis3718_himass}, 5. \cite{2019Leroystellarmass}, 6. \cite{2003HYPERLEDA_paturel}, 7. \cite{2014pak}, \cite{2015Zaritsky}, \cite{2023Cattorini}.}
\end{table}

\begin{table*}[hbt!]
    \centering
    \caption{UVIT observation details}
    \scalebox{0.8}{
    \label{tab:obs}
    \begin{tabular}{|l|l|l|l|l|l|l|l|}
        \hline
        {Object} & {RA} & {Dec} & {Observation ID} & {Observation date} & {P.I.} & {Available filters} & {Exposure times (s)} \\ \hline
        \multirow{2}{*}{}{NGC 3718} & \multirow{2}{*}{}{173.1452} & \multirow{2}{*}{}{53.067} & \multirow{2}{*}{}{G05 234T02 9000000} & \multirow{2}{*}{}{2016-05-29} & \multirow{2}{*}{}{Ashok Kumar Pati} & FUV (CaF2, BaF2) & 1102, 4857 \\
        &  &  &  &  &  & NUV (NUVB13, B4, N2) & 2175, 2895, 4959 \\ \hline
        \multirow{2}{*}{}{NGC 2685} & \multirow{2}{*}{}{133.894} & \multirow{2}{*}{}{58.734} & \multirow{2}{*}{}{A04 176T01 9000001} & \multirow{2}{*}{}{2017-12-21} & \multirow{2}{*}{}{Kshama S K} & FUV (CaF2) & 1442 \\
        &  &  &  &  &  & NUV (NUVB4, Silica15) & 2049, 6556 \\ \hline
        \multirow{2}{*}{}{NGC 4262} & \multirow{2}{*}{}{184.877} & \multirow{2}{*}{}{14.877} & \multirow{2}{*}{}{A02 058T02 9000001} & \multirow{2}{*}{}{2017-02-19} & \multirow{2}{*}{}{Omkar} & FUV (BaF2) & 1879 \\
        &  &  &  &  &  & NUV (Silica15) & 1890 \\ \hline
    \end{tabular}}
\end{table*}
\subsection{NGC 3718}
NGC 3718, also known as UGC 6524, Arp 214, and PRC D-18, is a member of the loose Ursa Major group, located at a distance of 18.7 Mpc \citep{1988Tully,1996Tully}. Its morphology is highly peculiar, characterised by a prominent dark dust lane that spans almost perpendicular and straight across the central bulge \citep{2005Krips}. Previous studies, such as \cite{1979Allsopp} and \cite{1985Schwarz},  have revealed a significant twist in the gas disc, which warps into an edge-on configuration where the straight dust lane is observed. The complex structure of the twisted HI disc of NGC 3718 is discussed in the studies of \cite{2004Pott}, \cite{2005Krips} and \cite{2009Sparke}. \cite{whitmore1990} compiled the first PRG catalogue, NGC 3718 was included in the “possibly related object” category (named as PRC D-18). Subsequent research by \cite{2009Sparke} confirmed that the innermost gas that orbits the stellar disc of the galaxy is nearly polar, whereas the outer gas rings are tilted at an angle of 30 to 40 degrees. 
%This observation further proved that NGC 3718 is a PRG.

\subsection{NGC 2685}
NGC 2685, also known as the Helix galaxy or the Spindle galaxy \citep{1961Sandage}, is located at a distance of 16.8 Mpc. \cite{whitmore1990} classified NGC 2685 as a "kinematically confirmed" PRG. The host galaxy exhibits a nearly perpendicular ring structure, with gas and stars orbiting close to the central body \citep{1997Eskridge2685,1998chenko2685}. 
The HI observations strongly suggest that the gaseous structure of NGC 2685 forms a coherent, extremely warped disc, with the appearance of two rings being attributed to projection effects. Also, at smaller radii, this disc is kinematically decoupled from the central stellar body  \citep{2009Jozsa}. NGC 2685 exhibits an outer component of neutral hydrogen. This outer component is positioned perpendicular to the polar ring and appears to align with the central body of the S0 galaxy \citep{2005Hagen2685,2009Jozsa}.

\subsection{NGC 4262}
NGC 4262 is a peculiar S0 galaxy in the Virgo cluster region and a classic example of a PRG, which is at a distance of $\sim$15 Mpc. \cite{whitmore1990} have first presented NGC 4262 as a kinematically related possible polar ring candidate. In the ACS VCS survey of 100 ETGs, \cite{Ferrarese2006} explored the morphology, isophotal parameters, and surface brightness profiles of NGC 4262. They investigated and explained that NGC 4262 has a thin dust filament structure, a small regular dusty disc, and a prominent stellar bar. Using GALEX observations of NGC 4262, \cite{Buson2011} detected an extended outer ring studded with UV-bright knots surrounding the galaxy body. This outer ring has an orthogonally inclined host galaxy. From these observations, \cite{moiseev2011MNRAS.418..244M}  added NGC 4262 in the new PRG catalogue as SPRC 33. Recently, \cite{2024akr_2} studied the galaxy using NGVS data and showed an optically faint ring structure that is orthogonal to the host galaxy. 

\begin{figure*}
    \begin{center}
        \includegraphics[width = 1.8\columnwidth]{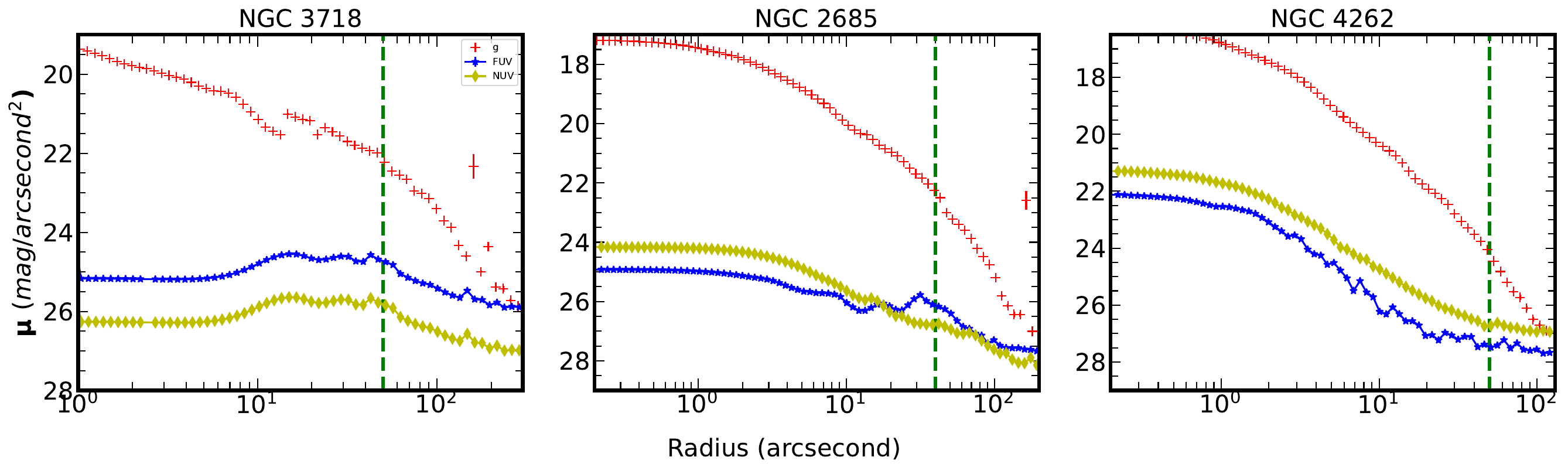}
        \caption{ The UV and optical surface brightness profiles of NGC 3718, NGC 2685 and NGC 4262. It is observed that the FUV profile dominates ($\sim$1 mag/arcsecond$^{2}$ than NUV) in NGC 3718, there is an overlap of FUV and NUV profiles in NGC 2685, and the NUV profile dominates in NGC 4262. The separation radius of the host galaxy obtained is indicated by a green vertical dotted line.}
        \label{fig:combo_image}
    \end{center}
\end{figure*}
\begin{figure}
\begin{center}
    \includegraphics[width = 0.9\columnwidth]{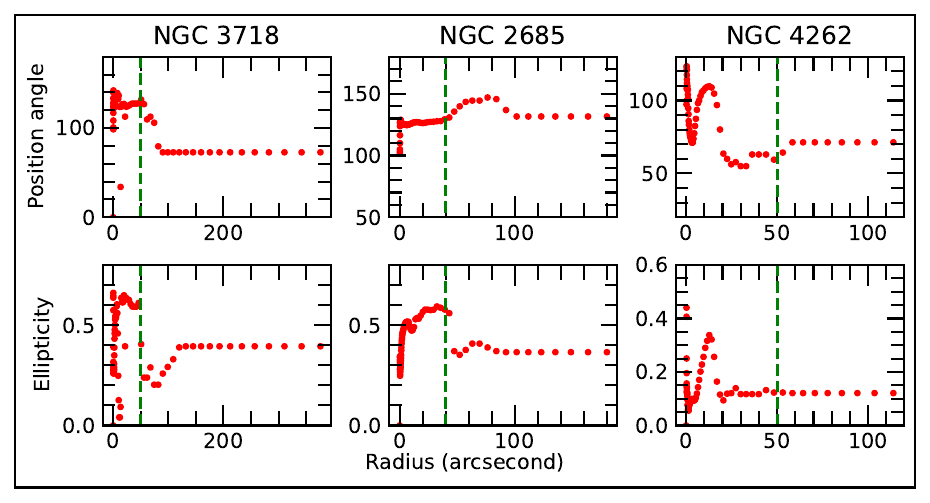}
    \caption{ Upper and lower panels show the variations of the optical isophotal parameters: position angle and ellipticity in SDSS $\it g$ band-filter as a function of radius. The separation radius of the host galaxy is indicated by a green vertical dotted line.}
    \label{fig:pa_ellp}
\end{center}
\end{figure}
\section{Data}
\label{sec:obs}
The observations of three galaxies presented in this paper include UVIT data from the AstroSat ISSDC data archive, optical data from Sloan Digital Sky Survey \cite[SDSS]{SDSS2000}, Near-infrared data from Two Micron AllSky Survey \cite[2MASS]{20062MASS} {and Mid-IR data from The Infrared Array Camera \cite[IRAC - Spitzer]{2004FazioIRAC,2004spitzer}}. UVIT, with a 28' field of view, observes simultaneously in FUV (130-180 nm), NUV (200-300 nm), and VIS (320-550 nm) filters. The spatial resolution of UVIT detectors is $\sim$\ang{;;1.5} with a field of view of 0.5 degrees and a pixel size of \ang{;;0.416} per pixel \citep{agrawal_uvit2006,kumar_uvit2012,Tandon_uvit2020}. 
The data reduction and calibration processes were conducted using the CCD Lab software \citep{postma_uvit2017}. The photometric zero points were taken from \cite{Tandon_uvit2020}. We followed the procedure outlined in \cite{2022UK} for the reduction and calibration of UVIT images. The observation log is listed in Table \ref{tab:obs}. The SDSS, 2MASS, and {IRAC} data were acquired from their respective public data archives (SDSS Science Archive and NASA/IPAC Infrared Science Archive). The Optical and UV colour composite images of the three PRGs are shown in Figure \ref{fig:color}.

\section{Analysis and Results}
\label{sec:anal_res}
In order to comprehend the evolutionary trajectory of the ring component within PRGs and their broader implications in the global context, we conducted surface photometry in both UV and optical images for three PRGs. Subsequently, SED analysis was performed on {the spatially resolved regions} of the galaxy. The exploration extended to identifying and examining the star-forming regions within the ring structure of the galaxies. Ultimately, the study sought to determine the position of these three galaxies within the overarching global evolutionary pathway.

\subsection{UV and Optical surface photometry}
\label{sec:UV_opt_phot}
We have carried out the surface photometry using Python elliptical isophotal analysis\footnote{https://photutils.readthedocs.io/en/stable/isophote.html} \citep{jedrzejwski1987} in the SDSS $\it g$ band, UVIT FUV, and NUV filters. During the fit, we allowed the isophotes free variation of the centre, ellipticity, and position angle (measured counterclockwise from the west). For NGC 3718, FUV is represented by the UVIT CaF2 filter, and the NUVB13 filter represents NUV. In the case of NGC 2685, FUV corresponds to the UVIT CaF2 filter, while NUV corresponds to the NUVB4 filter. Finally, for NGC 4262, FUV corresponds to the BaF2 filter, and NUV corresponds to the silica filter.\par

Based on the analysis and observations of optical and UV images, we have determined the size of the host galaxy component for three galaxies. The radial extent of the host galaxy component in these galaxies is determined based on the isophotal parameters' position angle and ellipticity. The abrupt changes in position angle and ellipticity are used to ascertain the radial extent of the host component in each galaxy \citep{2005Erwin_isophote,2011Li_isophote,2024akr}. Figure \ref{fig:pa_ellp} illustrates the ellipticity and position angle distribution for each galaxy. The host components of NGC 3718, NGC 2685, and NGC 4262 were determined to have radial extents of 50, 40, and {50} arcseconds, and the ring components extended up to 290, 200, and 130 arcseconds, respectively. Note that the radial extent values are derived directly from SDSS {\it g}-band observations, implying that a slight variation is possible from other studies \citep[e.g.][]{2009Sparke,2002Schinnerer,2024akr_2}. Hence, the ring and host components in the images have been distinguished. Subsequently, these specified radii are used to separate the ring and host components of the galaxies in this study from now on.
%, while Table \ref{tab:out_param} provides the output parameters for the extent of the ring and host.
    
The UV and optical surface brightness profiles for the sample of three PRGs are illustrated in Figure \ref{fig:combo_image}. From the profile of NGC 3718, it is revealed that there is $\sim$1 mag/arcsecond$^{2}$  dominance of FUV over NUV. It should be noted that in NGC 3718, the host component overlapped with a part of the ring, thereby slightly affecting the brightness distribution of the host component. In NGC 2685, we observe an overlap of FUV and NUV beyond \ang{;;20}. In the case of NGC 4262, NUV becomes more dominant ($\sim$0.9 mag/arcsecond$^{2}$) than FUV. The brightness profiles of NGC2685 and NGC 4262 given in Figure \ref{fig:combo_image} match with the UV observations reported in \citet{2017Rampazzo} and \cite{2018Bouquin}. When FUV emissions in a galaxy decrease, it indicates that OB stars are no longer forming. Also, FUV emission decreases more rapidly than NUV emission, suggesting an older population (compared with FUV) of stars contributing to the NUV. When we compare the results above, these variations in the UV profile are particularly observed beyond the radius of the host galaxy. This provides valuable information about the star formation history and stellar population of the ring component of the galaxies. 

\begin{table}
    \centering
    \caption{Filters and their wavelengths used in the SED fitting}
    \label{tab:filters}
    \scalebox{0.85}{
    \begin{tabular}{|l|l|c|l|l|c|}
        \hline
        {Telescope} & {Filter} & {$\lambda$ ($\AA$)} & {Telescope} & {Filter} & {$\lambda$ ($\AA$)} \\ \hline
        \multirow{7}{*}{UVIT} 
        & CaF2   & 1481  & \multirow{5}{*}{SDSS}  & u  & 3572  \\  
        & BaF2   & 1541  &                        & g  & 4750  \\  
        & B15    & 2196  &                        & r  & 6204  \\  
        & Silica & 2418  &                        & i  & 7519  \\  
        & B13    & 2477  &                        & z  & 8992  \\  \cline{4-6} 
        & B4     & 2632  & \multirow{3}{*}{2MASS} & J  & 12350 \\  
        & N2     & 2792  &                        & H  & 16620 \\ 
        &        &        &                        & K  & 21590 \\  \cline{4-6} 
        &        &       & \multirow{2}{*}{IRAC}  & 1      & 36000  \\
        &         &      &               & 2      & 45000   \\ \hline
    \end{tabular}}
\end{table}

\subsection{Spectral energy distribution of PRGs}
From the section \ref{sec:UV_opt_phot}, we successfully distinguished the ring and host components within the PRGs. In this section, we separated the ring and host components of these galaxies across various spectral regions, including UV, optical, and mid-IR wavelengths. This comprehensive approach allows us to thoroughly examine and understand the SED of ring and host components within these galaxies. 
\subsubsection{{Pre-processing and PSF degradation}}
{To obtain spatially resolved regions in the PRGs and perform SED fitting to study their physical properties, such as SFR, age, and mass, we processed the data in several steps. We removed foreground stars, estimated and subtracted the background, and matched the PSF of each image to the lowest resolution by degrading every band to the PSF of the 2MASS image. We used the proper motions of objects from Gaia Data Release 3 \citep[Gaia DR3;][]{GaiaCollaboration2023} to remove definite stars from the data \citep{2022Buzzosplus,2024akr_2}. In each photometric band, we used the photutils "MeanBackground" method \citep{larry_bradley_2023_Photutils} to estimate and subtract a local background for each star-masked image. For UVIT filters, given the minimal UV background, we manually estimated and subtracted the background. In order to standardise the resolution across datasets, we employed a Gaussian 2D kernel to convolve all images to the resolution (\ang{;;3.5}) of the 2MASS. Figure \ref{fig:all_filt} shows all the images of the three galaxies that were processed. The units and their error conversion procedures are executed in accordance with the guidelines stipulated in the \cite{2021pixedfit_abdurro}. }

%subsubsection{\st{Photometry of the ring and the host component of the galaxy}}

\begin{table}
    \caption{Models and parameter values used to build the SED. The values are derived from the SED fit and compared with the best available literature values. We use the default values for the parameters that are not listed here. References are: \cite{2019Boquien,Hunt2019,2021Turner_phangs}}
    \scalebox{0.85}{
    \begin{tabular}{|l|l|l|} \hline
     Module    & Parameter       & Value        \\ \hline
    \multirow{4}{*}{}{\begin{tabular}[c]{@{}l@{}}Star formation history\\ (sfhdelayedbq)\end{tabular}}            
    & tau\_main       & 0.1 - 13 Gyr               \\  
    & age\_main       & 0.1 - 13 Gyr                \\  
    & age\_bq      & 0.01,0.1,0.5,1 Gyr                \\ 
    & r\_sfr      & 0.01, 0.05,0.5,0.7,0.9               \\ \hline
            
    \multirow{2}{*}{}{\begin{tabular}[c]{@{}l@{}}Stellar population\\ (bc03)\end{tabular}}   & IMF & 0 (Salpeter)                        \\  
    & metallicity     & 0.02,0.05                      \\ \hline          
    \multirow{1}{*}{\begin{tabular}[c]{@{}l@{}}nebular\\ \end{tabular}}                       & logU              & -2    \\ \hline
    \multirow{3}{*}{}{\begin{tabular}[c]{@{}l@{}}Dust attenuation\\ (dustatt\_modified\_starburst)\end{tabular}} & E(B\_V) lines   & 0.01 - 1                \\  
    & E(B\_V) factor  & 0.44,0.25               \\  
    & powerlaw\_slope & -1, -0.1, -0.5, 0, 0.5, 1 \\ \hline
     \multirow{3}{*}{}{\begin{tabular}[c]{@{}l@{}}{Dust emission}\\ (dl2014)\end{tabular}} & qpah   & 0.47,1.77                \\  
    & umin  & 0.1,1,5             \\  
    & alpha & 1,2,3\\
    & gamma & 0.1,0.5,0.9 \\ \hline
    \end{tabular}}
    
    \label{tab:param}
\end{table}
\begin{figure*}
    \begin{center}
        \includegraphics[width = 2\columnwidth]{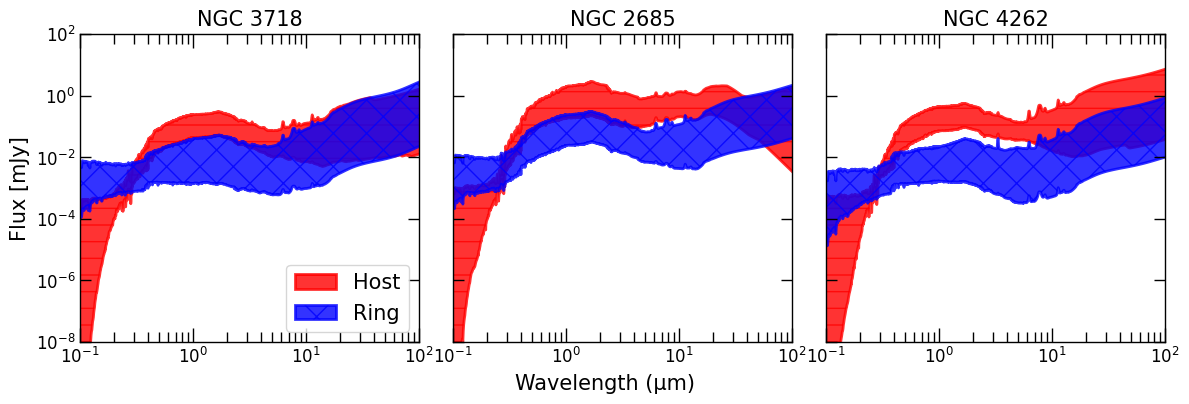}
        \caption{{SED of all the spatially resolved regions identified in galaxies NGC 3718, NGC 2685, and NGC 4262 from left to right. The blue and red shaded regions are based on the 16th and 84th percentiles of SEDs identified in the ring and host components, respectively. }}
        \label{fig:sed}
    \end{center}
\end{figure*}

\begin{figure*}
    \begin{center}
        \includegraphics[width = 1.8\columnwidth]{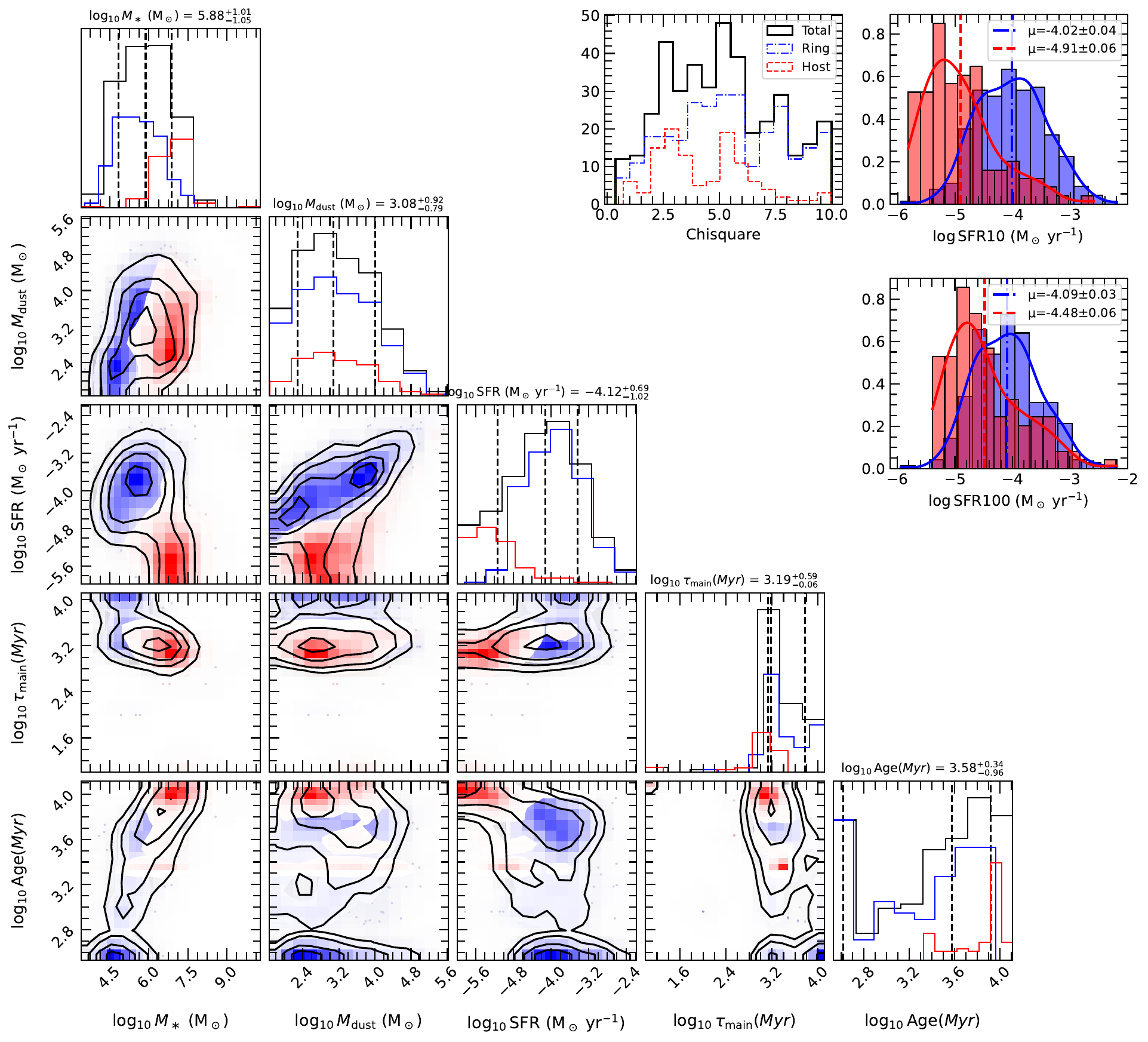}
        \caption{{The output parameters from SED fitting for the two-by-two combinations of parameters and histograms are presented for all four parameters: stellar mass, dust mass, SFR, e-folding time and age. The black contour regions represent the total stellar population distribution in all three PRGs. The mean values of each parameter are listed at the top of each histogram. Additionally, in each corresponding histogram, a list of three fractional quantiles [0.16, 0.5, 0.84] indicates the upper and lower errors. In the top right corner, the total best-fit chi-squared values, the SFR at 10 and 100 Myr are also shown. In all the plots, the red and blue colours represent regions belonging to the host and ring in the selected sample of PRGs.}}
        \label{fig:corner_plot}
    \end{center}
\end{figure*}
\subsubsection{Spatially resolved regions}
\label{sec:regions}
In our SED fitting analysis, we incorporated UVIT, SDSS, 2MASS, and IRAC photometric data. The specific filters utilised, and their corresponding wavelengths are given in Table \ref{tab:filters}. 
{In the case of spatially resolved regions in PRGs, we visually categorised them into three components: regions within the ring, the host galaxy, and regions where the host and the ring overlap. We used the ProFound package to identify the brightest regions in the NUV images, where NUV emission remains consistent across all these regions. ProFound \citep{2018Robotham,2018Robotham2}, an astronomical data processing tool in the R programming language, identifies peak flux regions within an image and delineates source segments using watershed deblending. We set a criterion that requires each detected region to cover a minimum of 10 pixels based on the resolution used during convolution. Then, these segments are iteratively expanded to encompass the full photometric profile \citep{2018Robotham}. The segmentation maps derived from the NUV images were overlaid on all other available filters to estimate the flux for each region.}

To identify the regions corresponding to the host, ring, and overlap areas, we used IR (IRAC 1) and UV (UVIT FUV) luminosity distributions as overlaid contours. Figure \ref{fig:con_reg} visually represents the selected regions within the host, ring, and overlap areas in the case of NGC 3718 (for NGC 2685, see Figure \ref{fig:con_2685}). For region selection, we applied criteria that require at least one filter from the UV, optical, and IR range to show a flux greater than zero. Additionally, the central regions of NGC 3718 and NGC 2685, which might be affected by possible AGN activity, were excluded as they fall within the overlap areas. This exclusion effectively mitigates AGN contamination in the selected regions. Finally, we obtained 407, 70, and 34 spatially resolved regions without any overlap in the central regions of the galaxies NGC 3718, NGC 2685, and NGC 4262, respectively.

\begin{figure*}
    \begin{center}
        \includegraphics[width = 2\columnwidth]{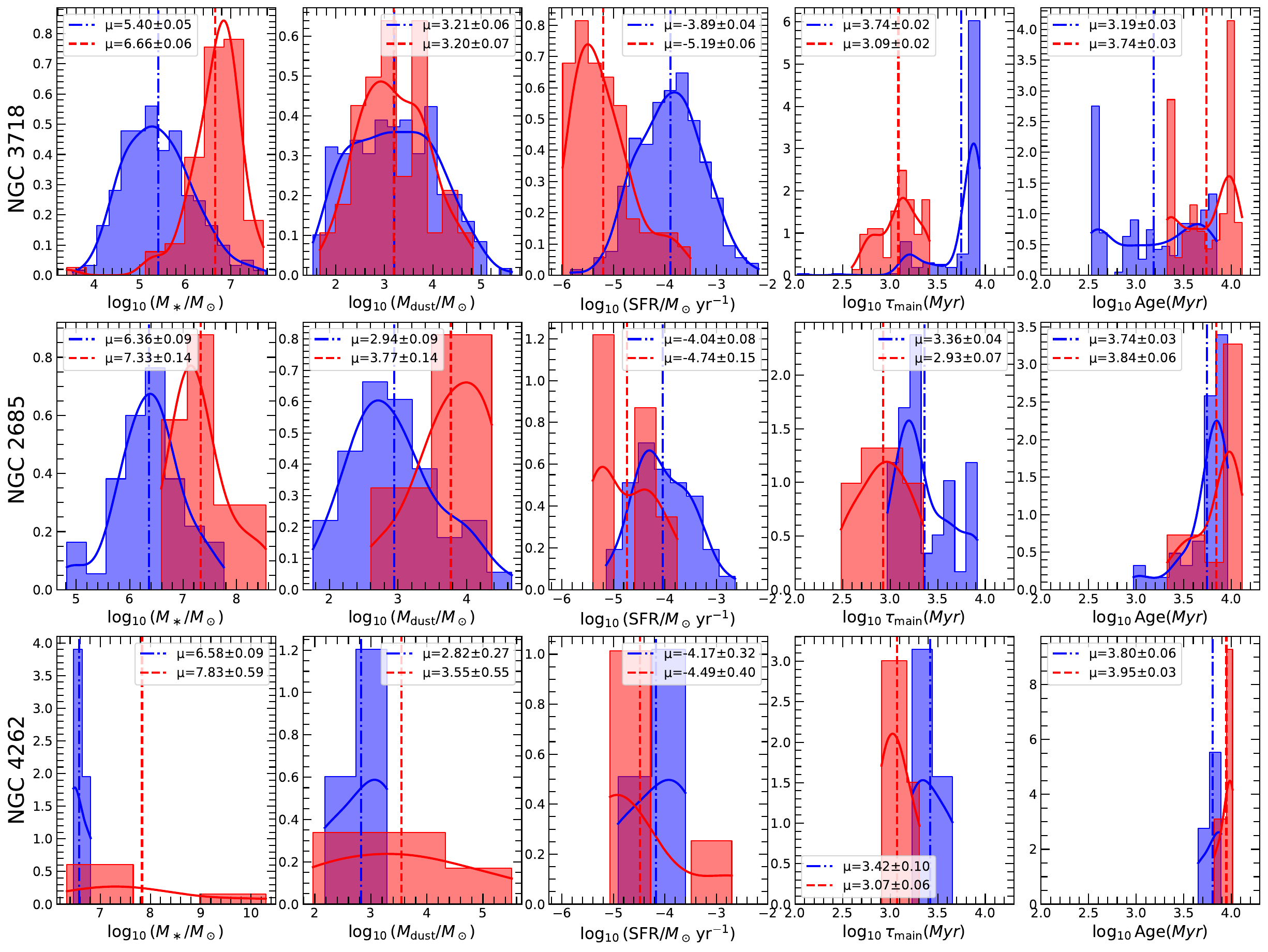}
        \caption{The histograms for the parameters—stellar mass, dust mass, SFR, e-folding time, and age—are presented separately for the host and ring regions, shown in blue and red, respectively. The first, second, and third rows correspond to the galaxies NGC 3718, NGC 2685, and NGC 4262. In each histogram, the mean values of the density distributions are labelled for both populations. These mean values represent the spatially resolved regions in the galaxies, not the total properties of the entire host and ring components.}
        \label{fig:hist_all_params}
    \end{center}
\end{figure*}

\subsubsection{SED modelling using CIGALE tool}
We used CIGALE\footnote{https://cigale.lam.fr/}, a Python-based tool that handles SED modelling and fitting. The SED modelling function allows us to construct comprehensive SEDs for the ring and host regions {}. This is achieved through the use of a combination of specialised modules designed to understand various characteristics. 

We employed the flux of each subcomponent across the wavelength range from FUV to Mid-IR (Table \ref{tab:filters}) and implemented models to calculate the optimal fit. {The images of all the filters of all three galaxies are shown in Figure \ref{fig:all_filt}.} For the star formation history, we utilised $\it{sfhdelayedbq}$ model \citep{2017Ciesla,2019Boquien}, incorporating an instantaneous recent variation in the SFR, either increase or decrease. Additionally, we utilised $\it{BC03}$ for stellar population modelling \citep{Bruzual2003}, $\it{Nebular}$ for nebular emission contribution \citep{2011Inoue} and $\it{dustatt powerlaw}$ for dust attenuation \citep{2019Boquien}. {We also utilised the dust emission model by \citet{Draine2014} to model the dust emissions within the galaxies,  which builds on earlier work by \citet{Draine2007}.  This model provides empirical templates for dust emission, dividing the emission into components associated with different radiation field intensities.}
Table \ref{tab:param} shows the models selected and the parameters chosen for computing every model in detail \citep{2019Boquien, Boquien2020}. \par

Using the models and parameters, we derived the SED for spatially resolved regions in the rings and host galaxies of all three galaxies, as shown in Figure \ref{fig:sed}. Notable differences emerge between the SEDs of the ring and host regions, particularly in the UV region. In NGC 3718 and NGC 2685, the UV flux in the ring regions is higher compared to the host regions. However, in the case of NGC 4262, compared to NGC 3718 and NGC 2685, the flux variation between the host and ring regions of NGC 4262 is less pronounced. \par
Due to the limited number of regions and to study the ring and host separately for all three galaxies together, we combined all the ring regions and host regions from the three galaxies. From the total regions identified across all three galaxies, we obtained approximately 70\% ring regions and 30\% host regions combined. The SED fits for our galaxies are deemed reliable, as their overall chi-square distributions have been thoroughly analysed. A small number of outliers ($\sim$2\%) with chi-square values exceeding 10 were excluded from the analysis to ensure robustness. The histogram in the top-right corner of Figure \ref{fig:corner_plot} illustrates the distribution of reduced chi-square values for the SED fits, combining both host and ring regions across the three PRGs. To examine the host and ring components together, we analysed key physical properties derived from the CIGALE SED fitting, including stellar mass, dust mass, SFR, e-folding time (\(\tau_{\text{main}}\) ), and age. Figure \ref{fig:corner_plot} presents corner plots for these parameters, highlighting the distinct physical properties of the host and ring regions. The distinct separation observed in these properties highlights significant differences in the stellar populations across these regions in all three galaxies. To quantify these differences, we performed both the Kolmogorov-Smirnov test and the permutation test. The p-values for stellar mass, SFR, \(\tau_{\text{main}}\), and age are extremely small (essentially zero), indicating significant differences between the ring and host components. In contrast, the p-value for dust mass ($\sim$0.4) suggests no significant difference. These findings emphasise the presence of distinct stellar populations in the ring and host components overall.
The disparity is particularly evident when comparing recent (SFR over 10 Myrs) and older (SFR over 100 Myrs) star formation rates, as shown in the top-right corner of Figure \ref{fig:corner_plot}. For the logarithmic SFR over 10 Myrs, reflecting the most recent star formation, the mean values for the ring and host regions are $-4.02 \pm 0.04$ and $-4.91 \pm 0.06$, respectively. In contrast, the logarithmic SFR over 100 Myrs, corresponding to earlier star formation, shows less variation in the ring regions ($-4.09 \pm 0.03$) but higher values for the host ($-4.48 \pm 0.06$). These observations suggest distinct stellar populations in the ring and host components. Moreover, the SFR in the ring regions remains relatively constant, declining rapidly in the host galaxies.

{Returning to our initial interpretation, we proposed that the ring components of PRGs evolve progressively in the sequence NGC 3718, NGC 2685 and NGC 4262. To further investigate, we analysed the physical properties of the host and ring regions separately for these galaxies. Since this study focuses on the resolved stellar population regions in the PRGs, we carefully selected regions to minimise contamination between the host and ring components, as discussed in section \ref{sec:regions}. Consequently, it is important to note that this study does not provide results representing the entire ring population or the complete properties of the host galaxies. Instead, the analysis reflects the characteristics of the spatially resolved regions that were specifically identified to ensure accurate SED fitting and comparative evaluation. Figure \ref{fig:hist_all_params} presents histograms of key parameters, including stellar mass, dust mass, SFR, \(\tau_{\text{main}}\), and age. The mean values obtained for each parameter from the SED analysis of resolved regions in three galaxies host and ring are listed in the table \ref{tab:cig_out}. The results reveal distinct stellar populations in the host and ring regions across all three galaxies. }
{Moreover, we observe that the difference in mean SFR values between the host and ring regions decreases from NGC 3718 to NGC 2685, with NGC 4262 showing the smallest difference ($\Delta \mathrm{SFR}_{\text{ring-host}} \sim1.4 > \sim0.7 > \sim0.2$, respectively). This supports the hypothesis of an evolutionary sequence among these galaxies, transitioning from NGC 3718 to NGC 4262. In terms of stellar mass, e-folding time, and age, the stellar populations exhibit significant differences between the host and ring regions in each galaxy. Interestingly, NGC 3718 shows no significant separation in dust mass between the ring and host components, suggesting that both regions contain similar amounts of dust. This might indicate heightened star formation activity in the ring and host regions.}
\begin{table*}
\centering
\caption{Mean values of the physical parameters for the spatially resolved ring and host regions of the three galaxies, derived from CIGALE SED fitting. These values represent only the spatially resolved regions and do not reflect the total properties of the host or ring components.}
\begin{tabular}{|l|l|l|l|l|l|l|}
\hline
Galaxy    & Component & $\log(\text{Age})$ [Myr] & $\log(M_\text{dust})$ [$M_\odot$] & $\log(M_\star)$ [$M_\odot$] & $\log(\text{SFR})$ [$M_\odot$/yr] & $\log(\tau_\text{main})$ [Myr] \\ \hline
NGC 3718  & Host      & $3.74 \pm 0.02$ & $3.20 \pm 0.06$ & $6.66 \pm 0.06$ & $-5.19 \pm 0.06$ & $3.09 \pm 0.02$ \\ 
          & Ring      & $3.19 \pm 0.03$ & $3.21 \pm 0.06$ & $5.40 \pm 0.05$ & $-3.89 \pm 0.04$ & $3.74 \pm 0.02$ \\ \hline
NGC 2685  & Host      & $3.84 \pm 0.06$ & $3.77 \pm 0.14$ & $7.33 \pm 0.14$ & $-4.74 \pm 0.15$ & $2.93 \pm 0.07$ \\ 
          & Ring      & $3.74 \pm 0.03$ & $2.94 \pm 0.09$ & $6.36 \pm 0.09$ & $-4.04 \pm 0.08$ & $3.36 \pm 0.04$ \\ \hline
NGC 4262  & Host      & $3.99 \pm 0.02$ & $3.44 \pm 0.27$ & $7.50 \pm 0.27$ & $-4.49 \pm 0.40$ & $3.01 \pm 0.06$ \\ 
          & Ring      & $3.80 \pm 0.06$ & $2.82 \pm 0.27$ & $6.58 \pm 0.09$ & $-4.17 \pm 0.32$ & $3.42 \pm 0.10$ \\ \hline
\end{tabular}
\label{tab:cig_out}
\end{table*}

{The \(\tau_{\text{main}}\) is the timescale over which the SFR declines exponentially, reducing to a factor of \(1/e\) of its initial value. A larger \(\tau_{\text{main}}\) indicates a slower decay of star formation \citep{2019Boquien}. This suggests that, in all three galaxies, the host is an older component with a star formation history that has gradually declined, while the ring is younger, likely more dynamically evolving, with star formation that has been sustained for longer and is now declining at a slower rate.}
The hosts consistently exhibit shorter \(\tau_{\text{main}}\) values (NGC 3718: \(3.09 \pm 0.02\), NGC 2685: \(2.93 \pm 0.07\), NGC 4262: \(3.01 \pm 0.06\)), indicating a rapid decline in star formation. Their older stellar populations (log(Age): 3.74–3.99) reflect earlier, more evolved star formation histories. In contrast, the rings show longer \(\tau_{\text{main}}\) values (NGC 3718: \(3.74 \pm 0.02\), NGC 2685: \(3.36 \pm 0.04\), NGC 4262: \(3.42 \pm 0.10\)), implying sustained star formation over longer periods. The younger stellar populations in the rings (log(Age): 3.19–3.80) suggest more recent formation, likely triggered by interactions or accretion events. This trend highlights a progression in which NGC 3718 represents an earlier stage, featuring a dynamically young ring and a relatively evolved host. NGC 2685 shows an intermediate state, where both components have similar ages, while NGC 4262 represents a later stage, with a fully evolved host and a gradually declining ring.

\begin{figure}
    \begin{center} 
        \includegraphics[width = \columnwidth]{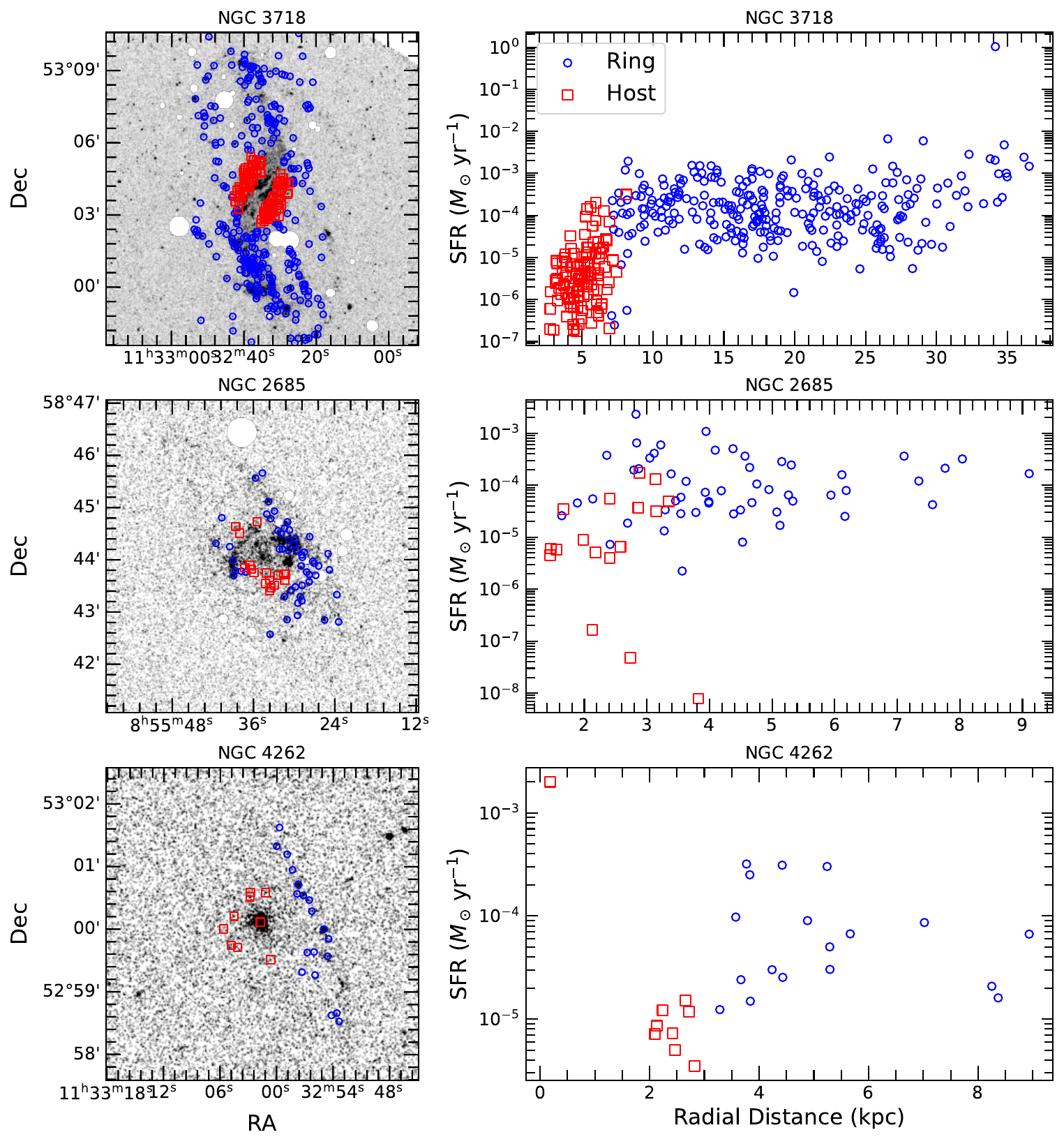}
        \caption{{The left panel shows the regions identified in the PRGs overlaid on the UVIT FUV images. Blue regions represent the regions in the ring, while red regions represent those within the host. The right panels illustrate the Radial distance versus SFR of the regions. We observe higher SFR in the ring component for NGC 3718 and NGC 2685 compared to NGC 4262.}}
        \label{fig:knots}
    \end{center}
\end{figure}

\subsection{Spatial and radial distribution}
Our comprehensive analysis shows that the trio of galaxies NGC 3718, NGC 2685, and NGC 4262 may trace distinct evolutionary phases within the PRGs. NGC 3718 suggests an initial stage, NGC 2685 occupies an intermediary position, and NGC 4262 represents the final stage of the evolution of the ring. To further investigate this progression, we conducted a detailed examination of the star-forming regions and UV properties at resolved scales within both the ring and host components of these galaxies.

Studying star-forming regions within a galaxy is crucial in enhancing our comprehension of its evolution. These regions possess distinctive characteristics, including size, SFR, radial distance, and distribution across morphological features, all of which contribute to our insights into the underlying mechanisms shaping the galaxy \citep{2022UK,2024robin}. A significant aspect of this analysis involves the utilisation of the intense FUV emissions primarily emitted by massive young {hotter O-type and B-type stars} to identify and track young massive star-forming regions in a galaxy \citep{kennicutt1998}.

{This study leveraged the advanced capabilities of multiple UVIT filters, which provide substantially higher spatial resolution compared to GALEX, enabling a more detailed examination of PRGs. From the results from the CIGALE SED fitting analysis, we investigated the resolved stellar population regions in PRGs at an unprecedented scale. The left side of Figure \ref{fig:knots} depicts the spatial distribution of identified regions within the three PRGs. The right panels of Figure \ref{fig:knots} illustrate the SFR of identified star-forming regions in the three PRGs as a function of radius. Our results reveal that NGC 3718 displays a higher number of star-forming regions in the ring structure than the host galaxy ($\sim$72\%). Furthermore, we observe that the SFR increases with distance from the centre in this galaxy.  Similarly, in NGC 3718 and NGC 2685, star formation is predominantly higher in the ring regions than in the host regions. In contrast, NGC 4262 shows comparatively lower star formation overall, with the host galaxy exhibiting a higher SFR than the ring structure. These observations highlight the distinct star-forming characteristics across the evolutionary stages of these PRGs, providing valuable insights into their star-formation processes.}

\subsection{The H I gas fraction versus NUV - r plane.}
\label{sec:HIgas}
\begin{figure}
    \begin{center}
        \includegraphics[width = \columnwidth]{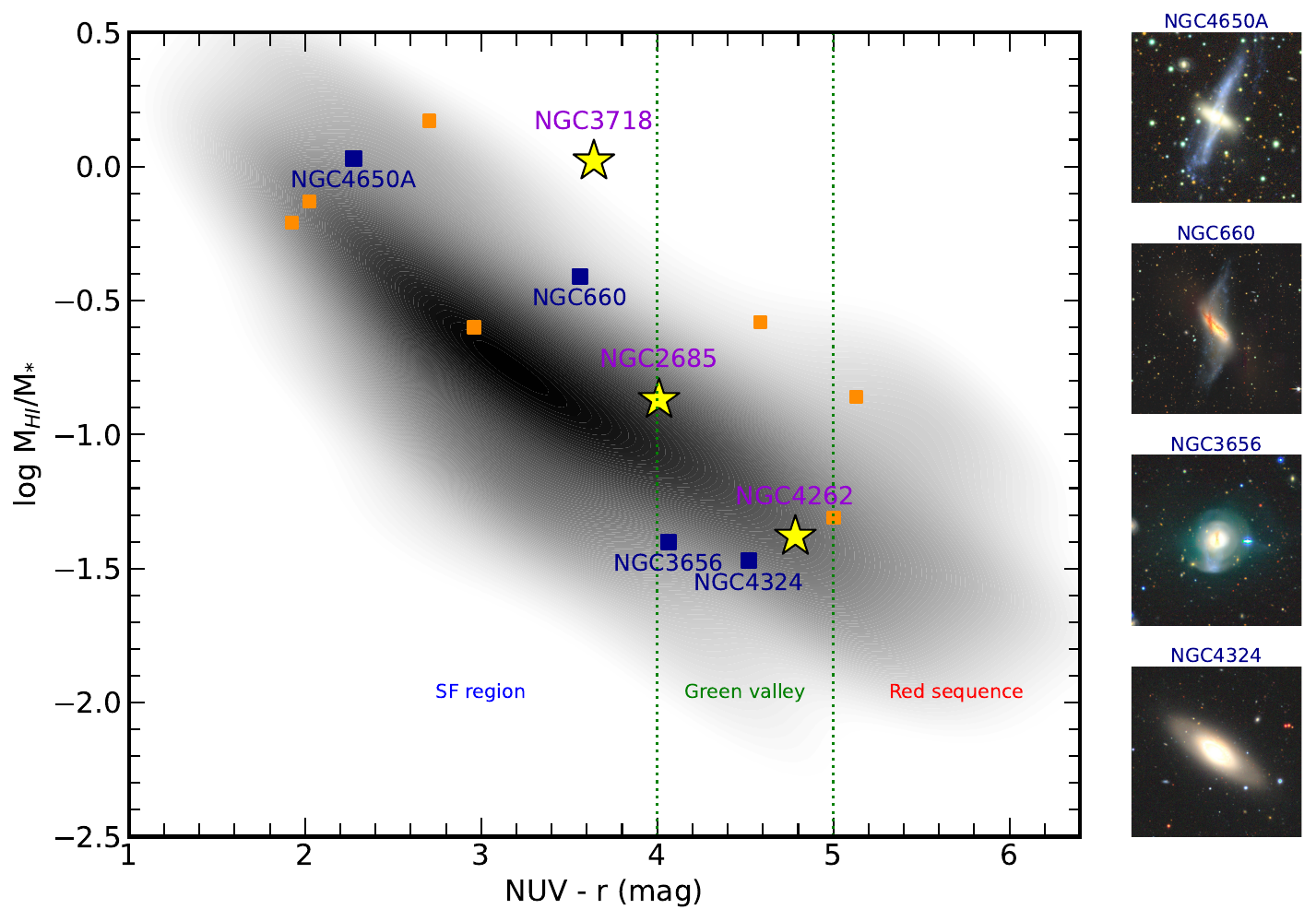}
        \caption{ The HI mass fraction as a function of NUV - r colour. The region within the green dashed line corresponds to the green valley from \citep{2014Salim}, and the region on the left and right correspond to the star-forming and quenched region, respectively. The black-shaded regions are the sample of galaxies from the GALEX Arecibo SDSS Survey \citep{2010Catinella,2018Catinella}. NGC 3718 lies within the star-forming region, NGC 2685 is transitioning into the green valley, and NGC 4262 is situated inside the green valley. The blue and orange squares represent the selected 11 sub-sample of PRGs from the literature. To facilitate a comparison of the ring structures, we included well-studied PRGs, specifically NGC 4650A, NGC 660, NGC 3656, and NGC 4324 (represented by blue-filled squares). Also, optical colour images from the DECaLS survey of these well-studied PRGs are displayed on the right side of the plot. These images illustrate various evolutionary phases in the morphology of the galaxies.}
        \label{fig:HINUV}
    \end{center}
\end{figure}

This section consolidates the preceding findings, offering a comprehensive overview of the evolutionary scenario of PRGs. The properties of HI gas and NUV-r colour in galaxies reveal their responsiveness to various physical conditions, providing valuable insights into the processes that drive the transition from blue, star-forming spirals to red, passively evolving ellipticals \citep{2009Cortese_Hughes,2010Catinella,2018Catinella}. Leveraging literature data, we have examined the HI gas mass fraction in relation to the NUV-r colour plane. In Figure \ref{fig:HINUV}, we present the positions of NGC 3718, NGC 2685, and NGC 4262 in the global evolutionary path of galaxies. For comparison, we present the sample of galaxies from the GALEX Arecibo SDSS Survey \citep{2010Catinella,2018Catinella} in the background, represented by a black-shaded region. To enhance comprehension, we have incorporated the Green Valley region from \cite{2014Salim} (Green dotted lines). The results align with our earlier observations, reinforcing the notion that NGC 3718 resides in the star-forming region, NGC 2685 is transitioning into the green valley stage, and NGC 4262 stands at the conclusion of the green valley phase, poised to evolve into a red sequence galaxy. These outcomes strongly support the claim that these three galaxies stand out as exemplary subjects for future investigations, providing crucial perspectives into the diverse evolutionary trajectories of PRGs. 
Additionally, we have selected a subset of 11 PRGs from the literature survey, all of which have HI observations, enabling us to observe their evolutionary phases and are represented in Figure \ref{fig:HINUV}. The details of the literature-selected galaxies and their properties are listed in Table \ref{tab:hi_prg}. Furthermore, for comparative analysis of morphological structures, we have included three optical images from the DECaLS survey on the right side of the plot, showcasing well-studied galaxies NGC 4560A, NGC 660, NGC 3656 and NGC 4324. Through these figures, it is evident that NGC 4650A exhibits a well-defined polar structure. According to this study, NGC 4650A may represent an earlier stage of PRG compared to NGC 3718. Subsequently, similar to NGC 3718, we observe NGC 660, and finally, akin to NGC 4262, we find NGC 3656 and NGC 4324.
\section{Discussion}
\label{sec:disc}

PRGs offer crucial insights into galaxy structure and evolution, ranging from understanding gas interaction processes to probing the shapes and distributions of their dark matter halos \citep{1996Reshetnikov3808,reshkinov1997,1998Bekki,2003Bournaud_Combes,2006macci,2009Sparke,2014Khoperskov,2020Quiroga,2021Khoperskov,2024Smirnov}. Since the initial discovery of potential stellar polar structures around nearby galaxies \citep{1961Sandage},  there has been a growing interest in studying these peculiar galaxies in detail. Previous studies, such as \cite{1998Bekki,2003Bournaud_Combes,2008ApJ...689..678B}, have investigated various formation scenarios for PRGs, including merging, accretion, and cold matter accretion. These studies also deliberate on the potential transformation of PRGs into E/S0-type galaxies. Simulations indicate that polar structures are not permanent, raising the possibility that rings could first form in HI, convert to optical structures, and eventually return to the galaxy host \citep{2003Bournaud_Combes,2021Khoperskov,2023Degwallaby}. This underscores the importance of studying the evolution of PRGs and their ring components.\par

This study identifies the possibility of NGC 3718, NGC 2685, and NGC 4262 as representative of three distinct stages in the evolutionary scenario of PRGs. Surface photometry in FUV and NUV from images provides valuable insights into the evolutionary processes of galaxies. The UV emission is highly sensitive to even subtle levels of star formation \citep{2011munoz_mateos,2017Rampazzo,2018Bouquin}. The UV surface photometry reveals distinct patterns in the surface brightness profiles of the sample PRGs, indicating variations in the dominance of FUV and NUV components in the ring structures beyond the radius of the host galaxies. The variation in the FUV luminosity in the ring components is directed towards the direction that NGC 3718 has dominant star formation and NGC 4262 has the least star formation among them. Analysing the UV part of the electromagnetic spectrum offers a straightforward approach to measuring the ongoing SFR in galaxies. {Examining resolved-scale regions within both the ring and host components adds a spatial dimension to the analysis. The identified regions exhibit varying levels of star formation, with NGC 3718 displaying higher SFR in the ring structures compared to NGC 2685 and NGC 4262.  The SED analysis further provides nuanced insights into their unique characteristics and evolutionary pathways (Figure \ref{fig:hist_all_params}).} Notably, the SED results support a consistent pattern across the three galaxies, suggesting a direct correlation between the evolving nature of the ring structure and the distinct evolutionary stages of PRGs. This observation supports the notion of distinct evolutionary phases within the PRGs, with NGC 3718 representing an initial stage, NGC 2685 occupying an intermediary position, and NGC 4262 indicating a more advanced stage of ring evolution.\par
NGC 3718 is a well-defined example of a galaxy with a central warp as mentioned by \cite{1985Schwarz}, \cite{2005Krips} and \cite{2009Sparke}. According to \cite{1994Arnaboldisparke} and \cite{2003Bournaud_Combes}, factors like inclination and mass distribution can influence the stability of PRGs and the likelihood of a warp occurring in the ring. In the case of NGC 3718, it is inferred that instability led to the initiation of a warp in the central region. This scenario provides a potential explanation for the second phase, as observed in NGC 2685. \cite{2009Jozsa} found that NGC 2685 exhibits two rings, one orthogonal to the host and another situated farther towards the centre, aligning with the host. Their findings suggest that this configuration could be an aftermath of a warp. However, it's crucial to acknowledge that the possibility of other scenarios influencing the formation of the ring structure cannot be dismissed. The integration of HI gas fraction and NUV-r colour plane (Figure  \ref{fig:HINUV}) analysis in this study aligns with the observed evolutionary stages, reinforcing the classification of NGC 3718 as a star-forming galaxy, NGC 2685 in transition to the green valley, and NGC 4262 positioned close to the termination of the green valley phase. In summary, the study provides a comprehensive narrative of the evolutionary progression within PRGs, combining observational data and modelling techniques to offer enlightening observations into the intricate processes shaping these unique galactic structures. The findings contribute to the broader understanding of galaxy evolution and highlight the potential of NGC 3718, NGC 2685, and NGC 4262 as exemplary subjects for future investigations in the field.

{ In addition to the three galaxies analysed in this study, a broader perspective on the evolution of PRGs requires examining a larger sample. In future work, we aim to expand our investigation to include a larger sample of galaxies, leveraging dedicated UVIT observations. The analysis of the three PRGs in this study serves as an initial exploration of the evolutionary scenario, while an extended study will enable a deeper investigation into the evolutionary stages of ring structures in PRGs. Such efforts will provide a more comprehensive view of the role of ring structures in galaxy evolution and contribute to the development of robust theoretical models for PRGs.}
\section{Summary}
\label{sec:conclu}
In this study, we examined the evolutionary trajectory of PRGs by utilising the most exemplary samples of galaxies: NGC 3718, NGC 2685, and NGC 4262. Our comprehensive investigation significantly enhanced our understanding of the star formation and star-forming activities within the ring components, along with shedding light on the evolutionary scenarios of PRGs. Employing SED fitting through CIGALE, we thoroughly explored the SED of the ring and host independently, marking the first instance of generating decomposed SED for PRGs. 
\begin{itemize}    
    \item The UV optical surface profiles show that the FUV profile dominates in NGC 3718, FUV overlaps NUV in NGC 2685, and NUV dominate in NGC 4262. This provides an initial insight into the star formation in the ring component for these galaxies, suggesting the existence of three distinct stages of ring evolution.
    
    \item {Using SED analysis, we studied stellar populations in the ring and host regions of the three PRGs. We observed notable differences in their physical parameters derived from the SED analysis, with the ring components generally being much younger than the host galaxies. The disparity decreases along the evolutionary sequence proposed from NGC 3718 to NGC 4262, indicating progressive galaxy evolution.}
    
    \item {A consistent pattern emerges when considering the SFR of the components. Our findings indicate that the star-forming regions within the ring component of NGC 3718 exhibit a higher SFR than the other galaxies. The SFR decreases progressively from NGC 3718 to NGC 2685 and then to NGC 4262. Additionally, the difference between the SFR of the host and the ring components approaches zero in the order of NGC 3718 to NGC 4262.}
    
    \item In the HI gas fraction versus the NUV - r plane, we observed that NGC 3718 lies within the star-forming region, NGC 2685 is transitioning into the green valley, and NGC 4262 is situated inside the green valley.
\end{itemize}
We suggest that the three galaxies are excellent examples for understanding the evolutionary pathway of PRGs: NGC 3718 is in the initial stage, followed by NGC 2685  (intermediate stage), and finally, NGC 4262 will be in the final stage of the evolution of PRGs.While these results give an initial insight regarding the class of PRGs and their evolution, they emphasise the need for a larger study. As a future endeavour, this study paves the way for modelling these evolutionary stages to facilitate further exploration.
%AKR and SSK want to acknowledge the financial support from CHRIST (Deemed to be University, Bangalore) through the SEED money project (No: SMSS-2220, 12/2022). We thank the Center for Research, CHRIST (Deemed to be University), for all their support during this work.
\paragraph{Acknowledgement}
{We would like to express our sincere gratitude to the referee for their insightful comments and constructive feedback, which significantly improved the quality and clarity of this article.} We thank the Center for Research, CHRIST (Deemed to be University), for all their support during this work. We thank Dr Smitha Subramanian, Dr Savithri H. Ezhikode, Dr Arun Roy, and Prajwel Joseph for their valuable comments on this work and wonderful discussions. We acknowledge the facility support from the DST FIST program (SR/FST/PS-I/2022/208). SSK and RT acknowledge the financial support from the Indian Space Research Organisation (ISRO) under the AstroSat archival data utilisation program (No. DS-2B- 3013(2)/6/2019). UK acknowledges the Department of Science and Technology (DST) for the INSPIRE FELLOWSHIP (IF180855). This publication uses UVIT, GALEX, SDSS, and 2MASS data. We gratefully thank all the individuals involved in the various teams for supporting the project from the early stages of the design and observations. This research has used the NASA/IPAC Extragalactic Database (NED), funded by the National Aeronautics and Space Administration and operated by the California Institute of Technology.  This work made use of  CCDLAB \citep{postma_uvit2017}, SAOImageDS9 \citep{2003Joyeds9}, TOPCAT \citep{2005Taylortopcat}, Astropy \citep{2013Astropy,2018Astopy}, Photutils \citep{larry_bradley_2023_Photutils}.

\paragraph{Data Availability Statement}
   All the UVIT data used in this paper are publicly available at \url{https://astrobrowse.issdc.gov.in/astro_archive/archive/Home.jsp}.  The data underlying this article will be shared with the corresponding author at a reasonable request. 
    
%paslike
\bibliographystyle{apalike}
\bibliography{example}
\begin{appendix}
\renewcommand{\thefigure}{\Alph{section}\arabic{figure}} 
\renewcommand{\thetable}{\Alph{section}\arabic{table}}
\setcounter{figure}{0} 
\setcounter{table}{0}

\section{Additional Information}
%The Figure \ref{fig:all_filt}
\begin{figure*}
    \begin{center} 
        \includegraphics[width=2\columnwidth]{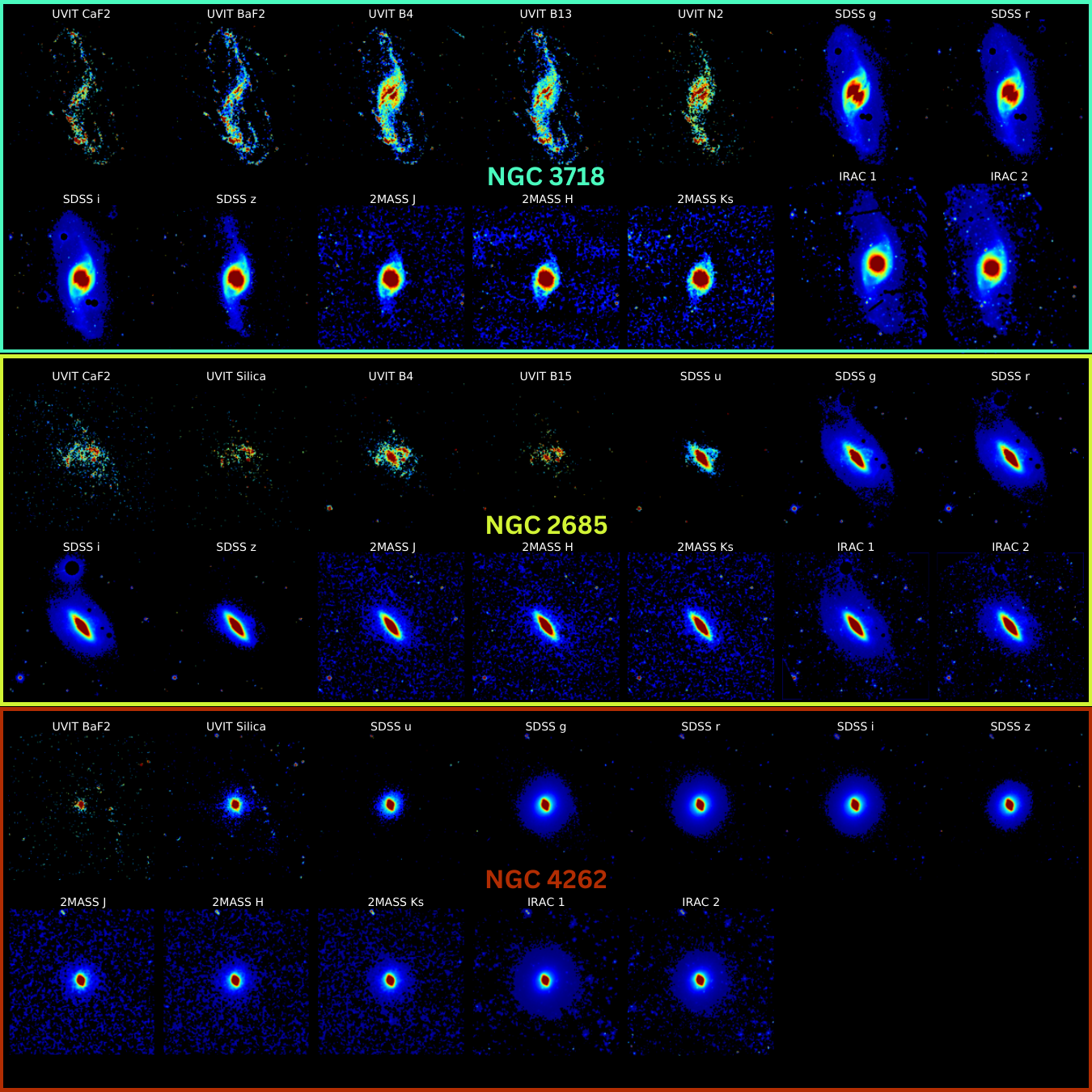}
        \caption{All available images of the three galaxies used in this study are shown here.}
        \label{fig:all_filt}
    \end{center}
\end{figure*}

\begin{figure*}
    \begin{center}
    \includegraphics[width = 1.5\columnwidth]{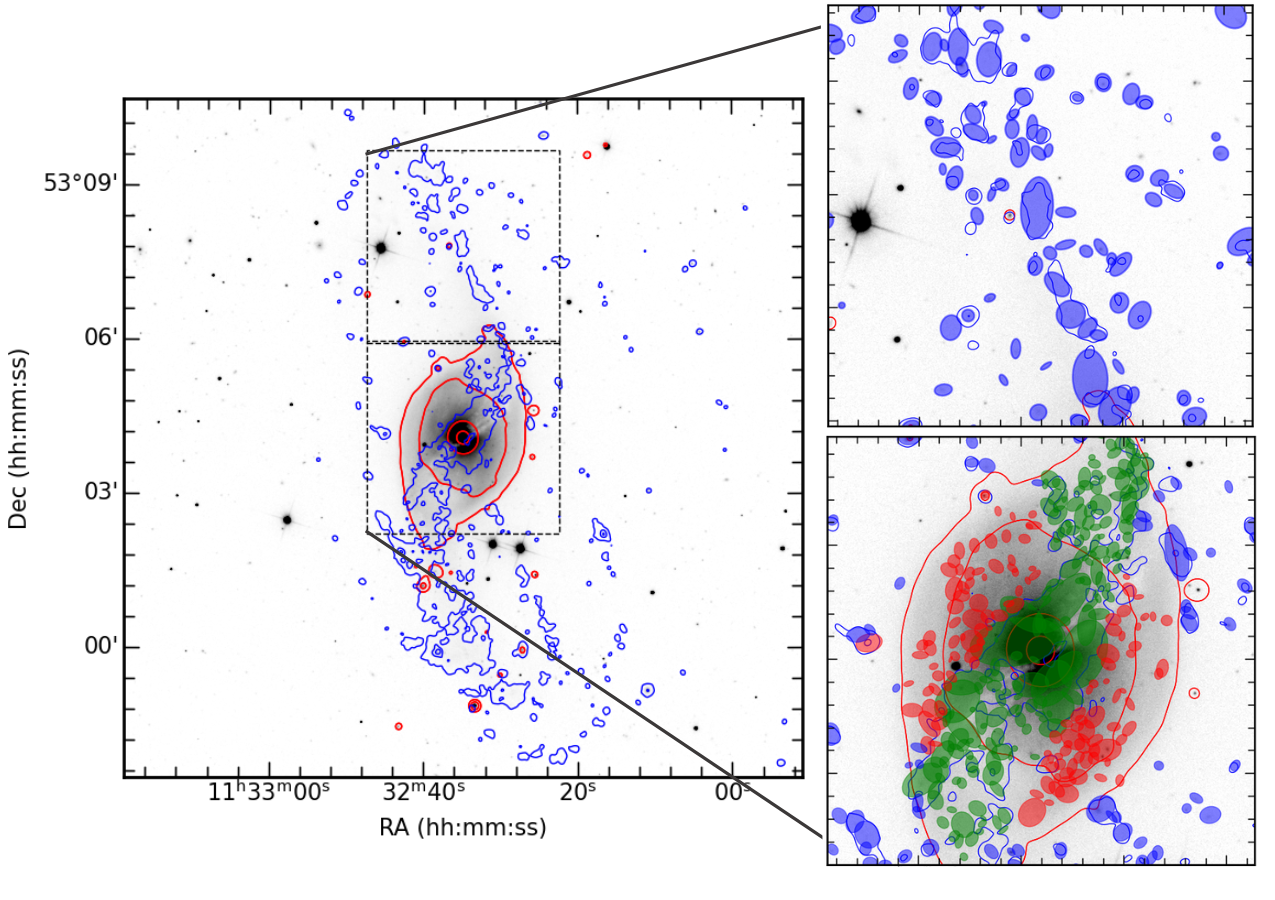}
    \caption{ The UV (blue) and IR (red) contours obtained from UVIT and IRAC images are overlaid on the optical SDSS image of PRG NGC 3718. On the right side of the image, the top panel shows a zoomed-in view of a ring component with identified regions marked by blue ellipses. Below this, a zoomed-in view of the centre of the galaxy displays identified regions in green (overlap) and red (host).}
    \label{fig:con_reg}
    \end{center}
    \end{figure*}

\begin{figure*}
    \begin{center}
    \includegraphics[width = 1.5\columnwidth]{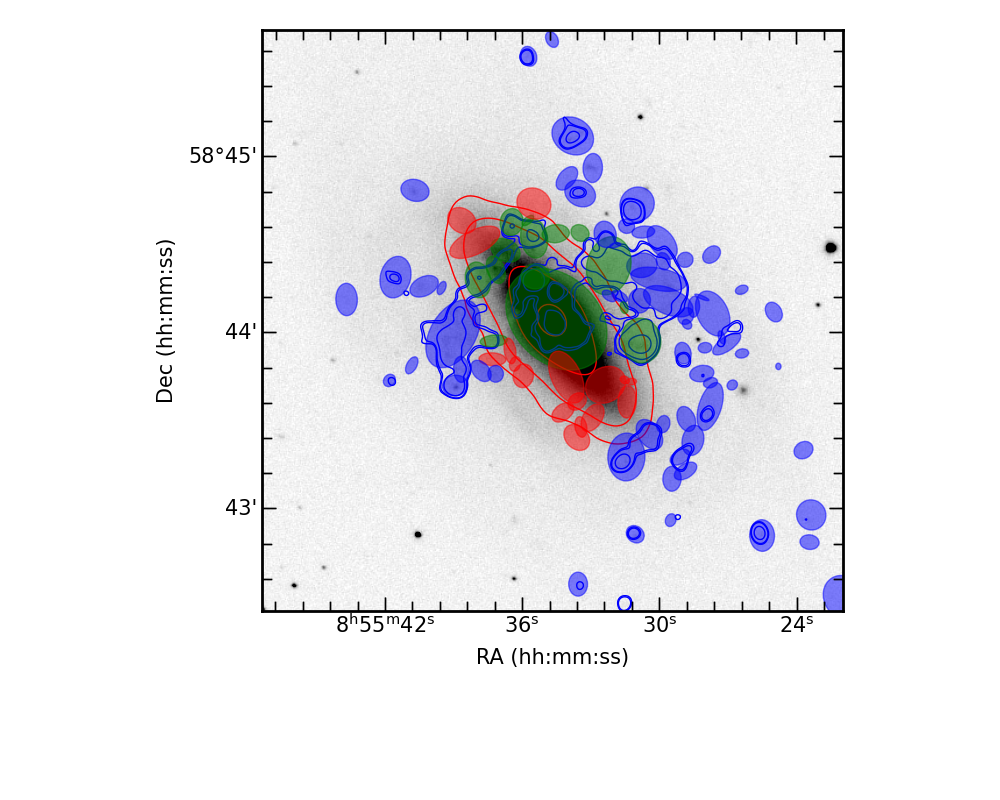}
    \caption{ The UV (blue) and IR (red) contours obtained from UVIT and IRAC images are overlaid on the optical SDSS image of PRG NGC 2685. The identified regions in green, blue and red represent overlap, ring and host regions, respectively.}
    \label{fig:con_2685}
    \end{center}
    \end{figure*}

\begin{table*}
    \caption{Properties of the selected subset of PRGs from the literature, including galaxy name, right ascension, declination, logarithmic neutral hydrogen mass, stellar mass, NUV, and r magnitudes. References are 1.\cite{2012Hall660}, 2.\cite{2018Parkash4650a} 3.\cite{2007Gildepaz},  4.\cite{2014Brown_UV_MIR}, 5.\cite{2012Seibert}, 6.\cite{2012Zacharias}, 7.\cite{2018Bouquin}, 8.\cite{2017Tempel}, 9.\cite{2011Bianchi}, 10.\cite{2015Alam}, 11.\cite{2013Gavazzi}}
    \label{tab:hi_prg}
    \centering
    \scalebox{1}{
    \begin{tabular}{|l|l|l|l|l|l|l|}
    \hline
    name     & RA (deg)  & Dec (deg)& logMHI ($M_\odot\ $) & logM* ($M_\odot\ $) & NUV (mag)& r (mag) \\ \hline
    NGC660   & 25.760   & 13.645  & 9.41$^{1}$   & 9.82$^{1}$   & 14.80$^{3}$  & 11.24$^{4}$ \\ \hline
    NGC4650A & 91.204  & -40.714 & 10.05$^{2}$  & 10.02$^{2}$ & 16.48$^{5}$  & 14.2$^{6}$    \\ \hline
    NGC4324  & 185.775 & 5.250  & 9.28$^{1}$   & 10.75$^{1}$ & 15.78$^{7}$ & 11.55$^{8}$  \\ \hline
    UGC05791 & 159.865 & 47.947  & 8.43$^{1}$   & 8.56$^{1}$  & 16.00$^{7}$  & 14.13$^{8}$  \\ \hline
    NGC3656  & 170.911 & 53.842  & 9.35$^{1}$   & 10.75$^{1}$ & 16.50$^{9}$  & 12.44$^{10}$   \\ \hline
    UGC09763 & 228.009 & 21.298  & 9.96$^{1}$   & 10.54$^{1}$ & 18.80$^{9}$   & 14.21$^{10}$   \\ \hline
    UGC09002 & 211.223 & 12.721  & 9.75$^{1}$   & 9.58$^{1}$  & 17.05$^{9}$  & 14.33$^{10}$  \\ \hline
    UGC04385 & 125.966 & 14.751  & 9.22$^{1}$   & 9.43$^{1}$  & 15.71$^{5}$  & 13.78$^{10}$  \\ \hline
    PRCD-51  & 229.310 & 21.585  & 9.88$^{1}$   & 10.74$^{1}$ & 20.52$^{9}$  & 15.38$^{10}$  \\ \hline
    UGC4332  & 124.907   & 21.114    & 9.65$^{1}$   & 10.96$^{1}$ & 18.39$^{9}$ & 12.96$^{11}$  \\ \hline
    IC51     & 11.600    & -13.442    & 8.84$^{2}$  & 9.44$^{2}$  & 15.80$^{9}$ & 12.84$^{10}$   \\ \hline
    \end{tabular}}
\end{table*}

As discussed in Section \ref{sec:HIgas}, we have used a set of sample galaxies from the literature to observe their evolutionary phases in the HI gas fraction vs. NUV-r plane relation.
\end{appendix}
\end{document}